\documentclass{kluwer}    

\newdisplay{guess}{Conjecture}

\begin{document} 
\begin{article}
\begin{opening}         
\title{Evidence for Solar-like Oscillations in Arcturus ($\alpha$ Boo)} 
\author{Alon \surname{Retter}, Timothy R. \surname{Bedding}}  
\runningauthor{Retter et al.}
\runningtitle{Solar-like Oscillations in Arcturus}
\institute{School of Physics, University of Sydney 2006, Australia}
\author{Derek \surname{Buzasi}}  
\institute{Department of Physics, 2354 Fairchild Drive, US Air Force Academy, 
CO 80840, USA}
\author{Hans \surname{Kjeldsen}}  
\institute{Theoretical Astrophysics Center, University of Aarhus, 8000 Aarhus
C, Denmark}
\date{August 7, 2002}

\begin{abstract}
Observations of the red giant Arcturus ($\bf \alpha$ Boo) obtained with the 
star tracker on the Wide Field Infrared Explorer ({\em WIRE}) satellite during 
a baseline of 19 successive days in 2000 July-August are analysed. The power 
spectrum has a significant excess of power at low-frequencies. The highest 
peak is at $\sim$4.1 $\mu$Hz, which is in agreement with the ground-based 
radial velocity and photometry study of Belmonte et al. (1990a; 1990b). The 
variability of Arcturus can be explained by sound waves, but it is not clear
whether these are coherent p-mode oscillations.
\end{abstract}

\end{opening}           

\section{Observations and analysis}  

After the failure of the main mission of the Wide Field Infrared Explorer 
({\em WIRE}) satellite, launched by NASA in 1999 March, its star tracker was 
used for asteroseismology of bright objects (Buzasi et al. 2000; Buzasi 2002). 
Here we report on preliminary analysis of observations of the red giant 
Arcturus ($\alpha$ Boo), which is the brightest star in the northern 
hemisphere, and has been observed in velocity several times (Belmonte et al. 
1990a; 1990b; Merline 1995). 

Nearly 1.2 million 0.1-s exposures were obtained during  19 successive days 
in 2000 July-August. Fig.~1 presents the light curve. The data from each 
orbit were binned into a single mean point with a typical error of about 15 
ppm. The light curve shows changes on time scales of hours to days. 

\begin{figure}
\vspace{9cm}  

\includegraphics{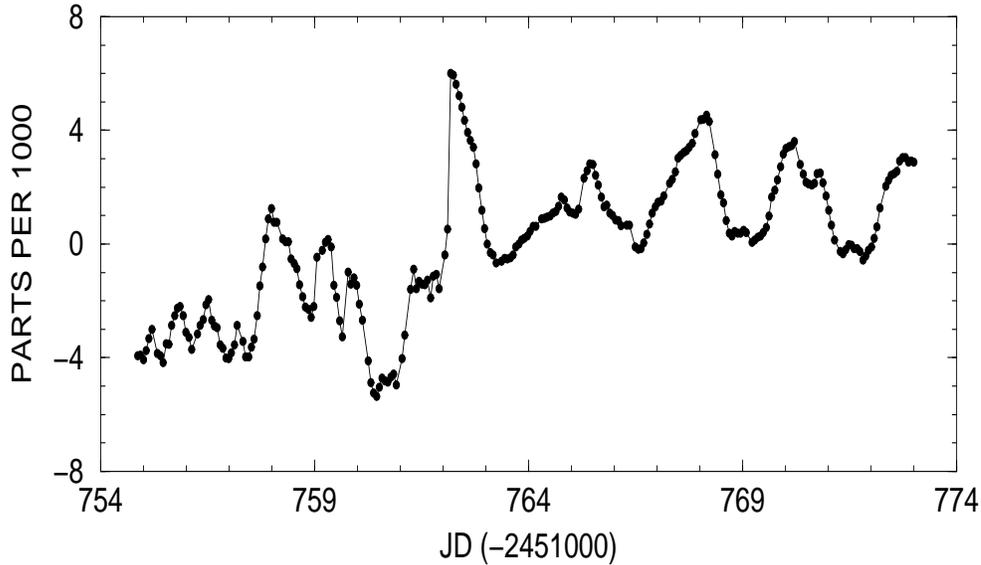}

\caption{The light curve of $\alpha$ Boo during the 19-d WIRE run in 2002. 
Each point represents a mean of about 4600 0.1-s exposures obtained 
during one 96-min orbit. The errors on a single point are smaller than 
its dimensions in the graph.}
\end{figure}

The upper panel of Fig.~2 displays the amplitude spectrum of the data. 
There is an excess of power at low frequencies. The highest peak at 
4.11 $\mu$Hz is in agreement with the ground-based radial velocity and 
photometry study of Belmonte et al. (1990a; 1990b).

\begin{figure}
\vspace{9cm}  

\includegraphics{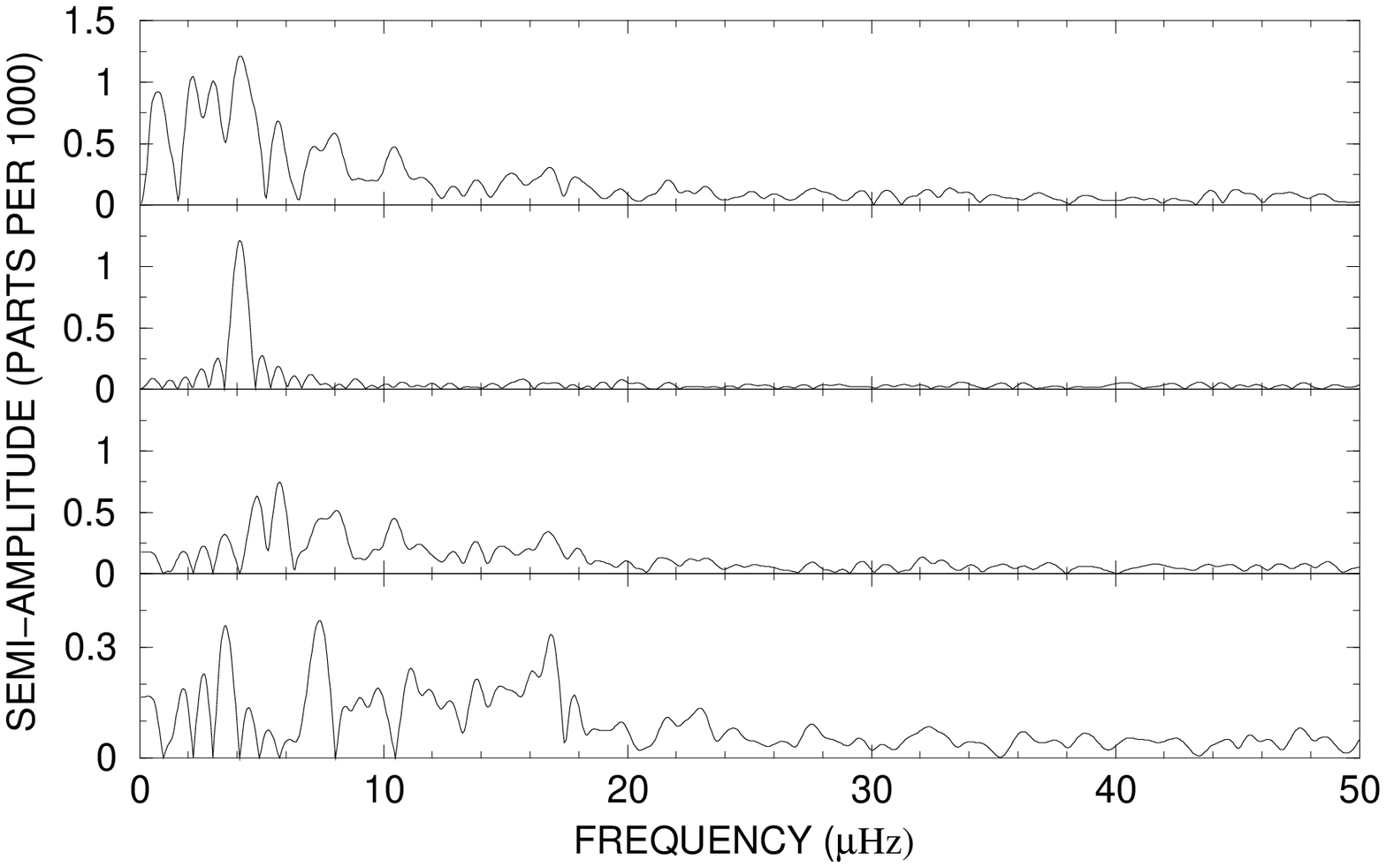}

\caption{Upper Panel -- Amplitude spectrum of the {\em WIRE} data.
There is a significant excess of power at low frequencies. The four highest
peaks are at 4.11, 2.26, 3.00 and 0.98 $\mu$Hz respectively.  Second panel
-- The spectral window (after planting a 4.11-$\mu$Hz sinusoid in the
window function with the same amplitude as in the data). Third panel -- The
power spectrum after prewhitening the four highest peaks. Lower panel -- after
prewhitening the eight highest peaks.}
\end{figure}

Table~1 lists the ten highest peaks in the power spectrum. The typical 
frequency error is $\sim$0.1 $\mu$Hz. Peak 4 (0.98 $\mu$Hz) is questionable 
as the data extend over only $\sim$2.5 cycles of it. Fig.~3 presents the
amplitude spectrum at the frequency range of 0-20 $\mu$Hz. The dotted 
vertical lines show the locations of the ten highest peaks.

\begin{table*}
\caption{The ten highest frequencies in the amplitude spectrum}

\begin{tabular}{rrccc}
     &           &                &              \\
Peak & Frequency & Semi-amplitude & comments     \\
     & ($\mu$Hz) & (ppm)          &              \\
     &           &                &              \\
1    & 4.11      & 980            &              \\
2    & 2.26      & 800            &              \\
3    & 3.00      & 760            &              \\
4    & 0.98      & 620            & questionable \\
5    & 5.73      & 520            &              \\
6    & 8.04      & 420            &              \\
7    & 4.92      & 380            &              \\
8    & 10.47     & 270            &              \\
9    & 7.35      & 270            &              \\
10   & 16.89     & 240            &              \\
\end{tabular}
\end{table*}

\section{Discussion}  

The amplitude and frequency of the power excess in Arcturus are consistent
with solar-like oscillations (Kjeldsen \& Bedding 1995).

The strongest peaks have a frequency spacing of 
$\Delta\nu\simeq$~0.8$\,\mu$Hz, which is in excellent agreement with the
expected value of 0.9$\pm$0.1 (Kjeldsen \& Bedding 1995), but not with the
value of 5.0 $\mu$Hz reported by Belmonte et al. (1990a; 1990b).  We note,
however, that our value of $\Delta\nu$ is very close to our frequency
resolution ($\sim$0.55 $\mu$Hz), which is set by the length of the run.
The same was true for the observations by Belmonte et al.

Could granulation noise, rather than oscillations, be responsible for the
excess power?  This is unlikely, since the variability is also detected in
velocity, with an amplitude of $\sim$60\,m/s at 4.3\,$\mu$Hz (Belmonte et
al. 1990a; 1990b).  The ratio of the photometric amplitude to the velocity
amplitude is as expected for sound waves, and about ten times greater than
expected for granulation noise.

We conclude that the variations in Arcturus can be explained by sound
waves, but it is not clear whether these are coherent p-mode oscillations
or perhaps a single mode with a short lifetime.  A longer time series is
required to distinguish between the two options.

\begin{figure}
\vspace{9cm}  

\includegraphics{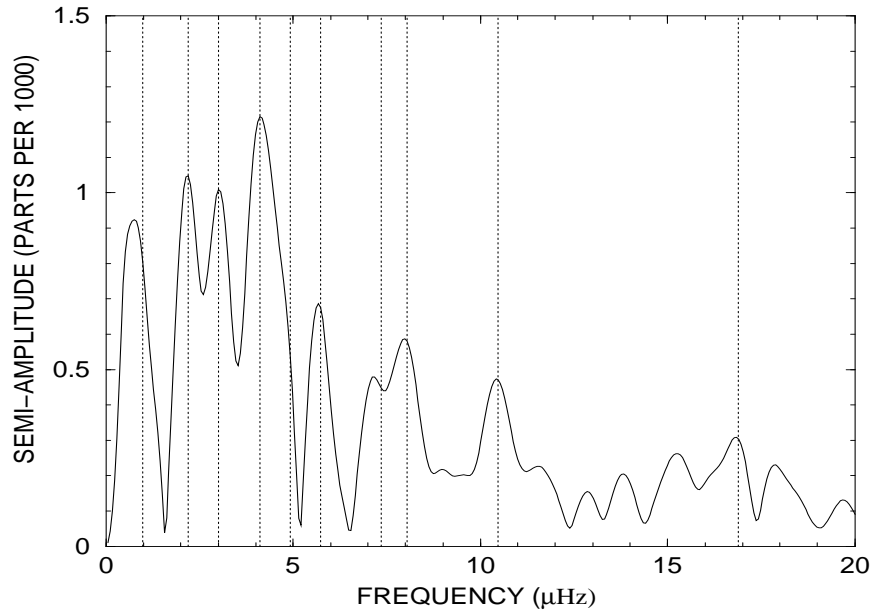}

\caption{Amplitude spectrum of the {\em WIRE} data in the range 0-20 
$\mu$Hz. The locations of the ten highest peaks are marked by vertical 
dotted lines.}
\end{figure}

\end{article}
\end{document}